\documentstyle[emulateapj,flushrt,epsf,rotate]{article}

\long\def\jumpover#1{{}}

\def\approxgt{\,\raise2pt \hbox{$>$}\kern-8pt\lower2.pt\hbox{$\sim$}\,}
\def\approxlt{\,\raise2pt \hbox{$<$}\kern-8pt\lower2.pt\hbox{$\sim$}\,}
\def \th{\thinspace}

\def \Teff{{$T_{\rm ef\!f} $}}
\def \Mo{{$M_\odot $}}
\def \Lo{{$L_\odot $}}

\def \eg{{{\it e.g.},\ }}
\def \etal{{\it et al.\ }}

\def \cf{{\it cf.\ }}

\def \ie{{{\it i.e.},\ }}

\def \viz{{\it viz.\ }}
\def \vs{{\it vs.\ }}

\def\K{\th K}
\def\sss#1{{{\scriptscriptstyle #1}}}

\def\sgn{{\rm sgn}}
\def\rn{\noindent\parshape 2 0truecm 8.8truecm 0.3truecm 8.5truecm}

\begin{document}

\submitted{\rm ASTROPHYSICAL JOURNAL, to be submitted}

\title{\textbf {Strange Cepheids and RR Lyrae}}  
\author{J. Robert Buchler$^{1}$, 
Zolt\'an Koll\'ath$^{2}$
}
 \begin{abstract}

Strange modes can occur in radiative classical Cepheids and RR~Lyrae
models.  These are vibrational modes that are trapped near the
surface as a result of a 'potential barrier' caused by the sharp
hydrogen partial ionization region.  Typically the modal number of
the strange mode falls between the 7th and 12th overtone, depending
on the astrophysical parameters of the equilibrium stellar models
(L, M, \Teff, X, Z).  Interestingly these modes can be {\sl linearly
unstable outside the usual instability strip}, in which case they
should be observable as new kinds of variable stars, 'strange
Cepheids' or 'strange RR~Lyrae' stars.  The present paper reexamines
the linear stability properties of the strange modes by taking into
account the effects of an isothermal atmosphere, and of turbulent
convection.  It is found that the linear vibrational instability of
the strange modes is resistant to both of these effects.  Nonlinear
hydrodynamic calculations indicate that the pulsation amplitude of
these modes is likely to saturate at the millimagnitude level.
These modes should therefore be detectable albeit not without
effort.

\end{abstract}

 \keywords{oscillations of stars - \th Cepheids - \th s--Cepheids}

{\bigskip
        {\footnotesize
$^1$Physics Department, University of Florida, Gainesville, FL, USA
 
$^2$Konkoly Observatory, Budapest, HUNGARY
}}

 \section{Introduction}

Strange modes were discovered and named by Saio, Wheeler \& Cox
(198) in highly nonadiabatic stars. (For a general reference \cf
Gautschy \& Saio 1996).  In a recent paper (Buchler, Yecko \&
Koll\'ath 1997; hereafter Paper I) it was demonstrated that strange, or
surface modes not only exist also in classical Cepheids, thrive
there.  Paper I showed that the strange modes are vibrational
modes that are predominantly surface modes that exist quite
naturally even in the adiabatic limit.

This was exhibited with the help of simple transformations of the
radial coordinate to an acoustic depth $z$, defined through $dz =
dr/c_s$ and of the radial displacement vector $\psi = \sqrt{\rho
c_s} r \delta r$, where $c_s$ is the sound speed, with which the
{\sl adiabatic} radial pulsation problem for radiative models can be
reduced to a time-independent Schr\"odinger equation
 \begin{equation}
  -{d^2\psi\over dz^2} +V(z) \psi = \omega^2 \psi
 \label{eq_psi}
 \end{equation}
The potential
 \begin{equation}
 V(z) = {1\over \sqrt{\rho c_s/r^2}}  {d^2 \over dz^2}\sqrt{\rho c_s/r^2}
 -{4Gm\over r^3}
 \label{eq_pot}
 \end{equation}
 incorporates thus the spatial variations of density $\rho$, sound
speed $c_s$ and spherical effects, as well as gravity.  This problem
is very similar to that encountered in the study of wind
instruments, such as the trumpet, where Eq.~1 is known as the
Bernoulli-Webster equation.

In Cepheids and RR~Lyrae the hydrogen partial ionization, in which
$c_s$ varies extremely rapidly, produces a very sharp and enormously
high potential barrier, that splits the pulsating envelope into two
almost disjoint parts (\cf Paper I).  Using the scaled eigenvectors
$\Psi$ one readily sees that the 'normal' vibrational modes have a
large amplitude in the interior region.  It is however also possible
for some modes to be predominantly trapped in the outer region --
these surface modes are the 'strange' modes.

 When, as is customary, the modes are classified with increasing
number of nodes (and thus increasing frequency) the strange modes
appear when a quarter wavelength fits approximately into the surface
region.  For a typical 5\Mo\ Cepheid on the left of the instability
strip one finds that the strange modes have $n$ = 10 -- 12, and
somewhat smaller $n$ on the cold side of the instability strip
because there is more matter outside the partial ionization region
of H.

When nonadiabatic effects, which are weak in Cepheids, are properly
taken into account, the modal frequencies acquire of course an
imaginary component.  Most of the stellar envelope is damping, but
the partial ionization zones cause strong pulsational driving.
Since the strange mode has at best a very small amplitude in the He
partial ionization region it thus experiences no driving or damping
there.  All its driving must occur in the H partial ionization
region.  It is not possible to extend the simplified
description of driving found in textbooks (\eg Cox \& Giuli 1965)
to modes higher than the fundamental or the first overtone.  For the
higher modes the driving or damping are a delicate function of the
relative phase of the pressure and specific volume eigenfunctions
which exhibit an increasing number of oscillations (related to the
nodes in the adiabatic limit) (\eg Glasner \& Buchler 1993, Buchler,
Yecko, Koll\'ath 1997) as the work-integral shows
 \begin{eqnarray}
  W_k &=& {2\pi \over I_k} \th
  {\rm Im} \int \delta p_k^*\th\delta v_k \th\th dm   \\ 
      &=& - {2\pi \over I_k} 
         \int |\delta p_k|\th|\delta v_k| \sin (\phi_v-\phi_p)_k \th\th dm\\
  I_k &=& \omega_k^2 \int |\delta r_k|^2 \th\th dm
 \label{eq_work}
 \end{eqnarray}

\centerline{{\vbox{\epsfxsize=10cm\epsfbox{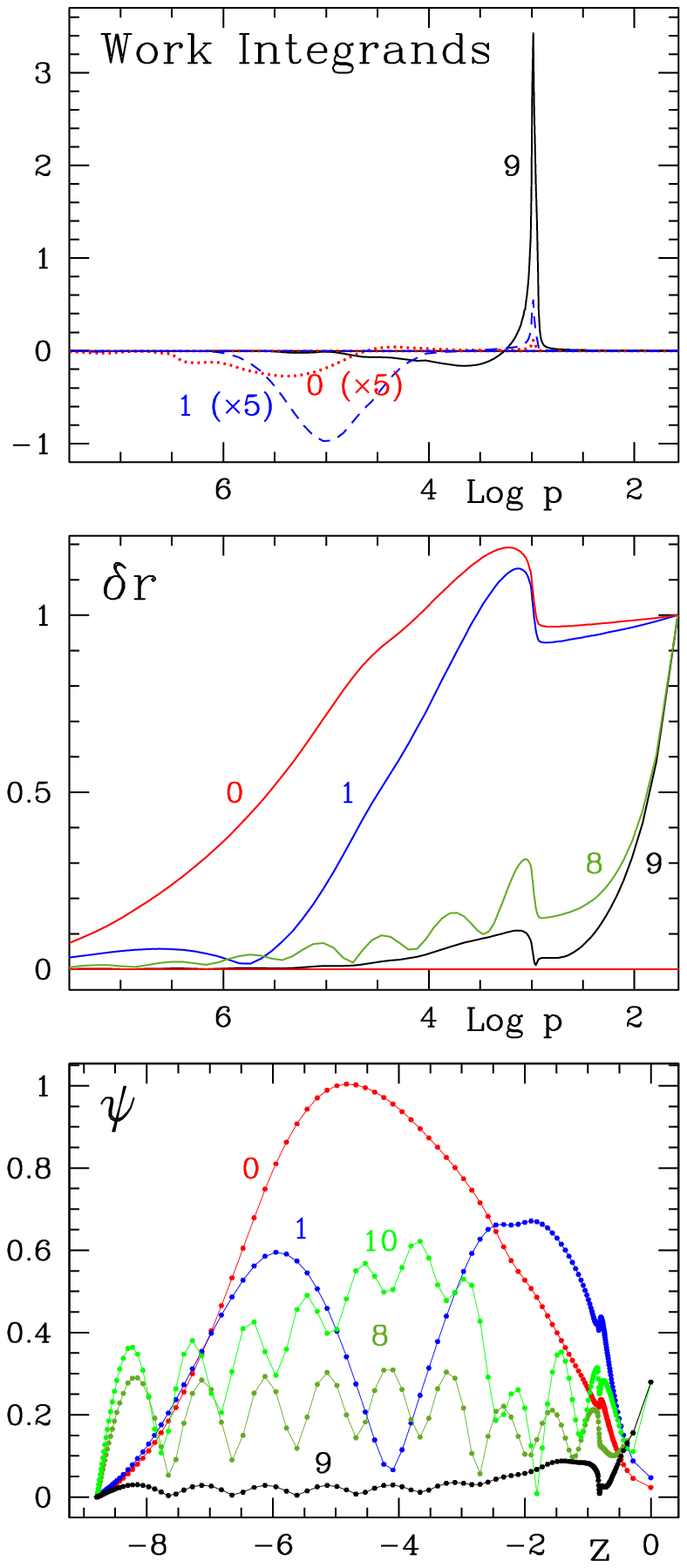}}}}
\noindent{\small Fig.~1. -- 
Cepheid model, M=8, L=10000, \Teff=6700\K\ model; {\sl Top:}
work-integrand  w \vs
Log$_{10}$(pressure) [cgs] for modes 0 (fundamental), 1 (first
overtone) and 9 (strange mode); 
{\sl middle:} radial displacement eigenvectors for modes 0, 1, 8
and 9; {\sl bottom:} transfored radial displacement eigenvector
($\Psi$) as a function of acoustic depth z (for clarity mode 1 has been scaled
up by a factor of 2 and modes 8 --10 by 12). 
}
\vskip 3mm

In Fig.~1 (middle) we exhibit the linear displacement eigenvectors
(modulus) as a function of Log p of the fundamental, first overtone
and strange 9th overtone for a Cepheid model with M=8\Mo,
L=10\th000\Lo, \Teff =6700\K, X=0.70, Z=0.02.  Clearly that the
strange mode is very strongly restricted to the surface region.

This even more apparent in the bottom of Fig.~1 which displays the
scaled displacement eigenvectors $\psi$ (\cf Eq.~1) as a function of
acoustical depth $z$ for the fundamental, first overtone, and for
overtones 8 through 10.  As pointed out in Paper I, this
representation of the modes as $\psi(z)$ brings out the vibrating
string nature of the modes, and shows that the strange mode (9th
overtone) is mainly a surface mode in contrast to its neighbors.

In the top Fig.~1 which displays the work-integrand as a function of
Log p, the integrand is normalized so that the area under the curve
represents the relative growth-rate $\eta = 2 \kappa_k \th P_k$. The
solid line represents the strange mode (9th overtone).  For
comparison we also show the work-integrands (multiplied by 5 for
improved visibility) for the fundamental (dotted) and first overtone
(dashed).  The latter modes are very stable for this model which is
on the left of the instability strip (\cf Fig.~1).  The strange mode
is restricted to the surface regions and its driving comes entirely
from the H partial ionization region around Log p $\sim 3$.

 Some simple {\sl a posteriori} explanation of the odd stability
properties of the strange modes can be given.
The strange modes penetrate much less deeply than the normal modes.
For the strange modes $W_k$ is thus (a) totally dominated by the
outer regions, in particular the hydrogen partial ionization region,
and (b) $W_k$ is much smaller.  However, its the moment of inertia
$I_k$
is also very small because the mode is localized in the surface.
Since the smaller work-integrals $W_k$ are normalized by the smaller
$I_k$, the strange modes still have relative growth-rates that are
comparable in magnitude to those of the normal modes.  However,
because only hydrogen driving contributes, the growth-rates of the
strange modes are egregious as was already shown in Paper I.  In
that paper strange modes were found to be vibrationally {\sl
unstable} to the left (blue) of the usual fundamental and first
overtone instability strips.

Furthermore nonlinear hydrodynamical modelling indicated that the
strange modes can give limit cycles, albeit with rather small
pulsation amplitudes, typically in milli-magnitudes.  This has
opened the exciting possibility of observing 'strange Cepheids', and
of using such additional information to put constraints on
theoretical Cepheid modelling.

Two questions were left unanswered in Paper~I, however.  The first
concerned the surface boundary condition, namely the disregard of
the surface impedance, due to the reaction of the isothermal
atmosphere and of running wave or acoustic damping.  The second
resulted from the neglect of turbulence and convection which can
also affect driving and damping.  Here, in \S2 we reexamine and
improve on the outer boundary condition.  In \S3 and \S4 we examine
turbulent convective Cepheid and RR Lyrae models.  In \S5 we present
nonlinear pulsations of turbulent convective models and we conclude
in \S6.

\section{Isothermal Atmosphere and Boundary Conditions} 

In paper I the models were integrated up to a point where the gas
pressure was small, typically a few times the radiation pressure.
In the linearization of the discretized equations we imposed a
standard condition of a zero Lagrangean pressure variation, $\delta
p_*=0$, \ie to an open and perfectly reflecting boundary.  This
boundary condition disregards two effects, first, running, escaping
waves, and second, inertial effects due to the isothermal
atmosphere.  In acoustic parlance, these effects give rise to
resistive and reactive surface loads (Morse \& Ingard 1968).

 It is interesting to note that the change of variables from $\delta
r$ to $\Psi$, because of the density dependence, effectively changes
the problem from an 'open quarter-wave length tube' to a 'closed
half-wave length tube'. (Here we have taken the stellar surface to
be defined as the point where the gas pressure equals two times the
radiation pressure, \ie at a very low density.)

Another way of looking at this is to say that when $\omega$ is
sufficiently large, the concomitant wave length gets sufficiently
short compared to the the isothermal scale height $1/\alpha$ so that
reflection at the surface becomes incomplete and acoustic energy is
lost into the atmosphere.

\vskip 1cm
\centerline{{\vbox{\epsfxsize=10cm\epsfbox{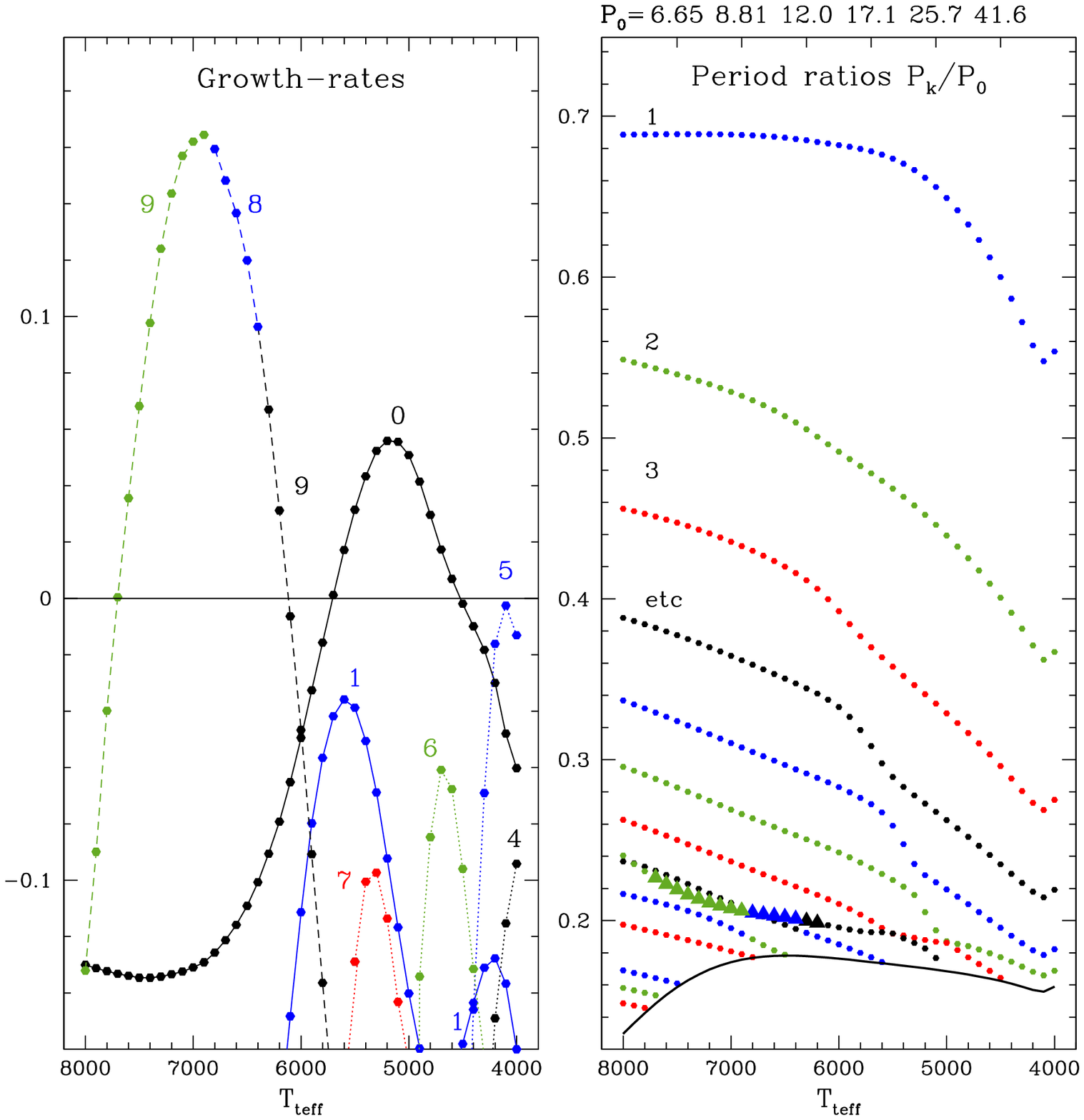}}}}
\noindent{\small Fig.~2. -- Cepheid model, M=8\Mo, L=10000\Lo.
{\sl Left}: growth-rates as a function of \Teff, positive growth-rates
correspond to linear unstable modes.  
The modes are labelled with the modal number.
{\sl Right:} period ratios
$P_k/P_0$ as a function of \Teff.  The fundamental period $P_0$ [d] 
is indicated on top.  
The triangles denote the unstable overtone modes. 
The solid line at the bottom is the isothermal cut-off period ratio.}

\vskip 10pt

In the optically thin outer region of the Cepheid model is
essentially isothermal $T\rightarrow (1/2)^\sss{1/4}$ \Teff \th\th.
To a good approximation the density decreases exponentially with
scale height $\alpha^{-1} = {\cal R} T/(\mu g)$, where $g= GM/r^2$
is essentially constant.  This gives rise to a potential barrier
which rapidly flattens to a constant value of $V_* = (c_s \alpha/
2)^2$.  In Buchler, Whiting \& Kollath (2001) it is shown that the
presence of the isothermal atmosphere can be approximately taken
into account through a boundary condition that is imposed at the
base of this atmosphere for the modes below the barrier ($\omega <
\omega_c$).  The modes above the barrier disappear into the
continuum and are no longer astrophysically relevant.
\begin{equation}
{\cal S} \equiv 
     (\Psi'/\Psi)_o = -i k {(\omega+k) +(\omega -k) \exp(-2 i k \Delta)
           \over (\omega+k) -(\omega-k) \exp(-2 i k \Delta)}
\end{equation}
where $k \equiv  k_r +i \th k_i = \sqrt {\omega^2-V_*}$.
When $|k_i \Delta|$ is large then ${\cal S} \sim i k \th \sgn\th k_i $.
The approximation that is inherent in the expression for the
boundary condition is that the potential is a square barrier of
width $\Delta$.  This translates into a boundary condition
 \begin{equation}
 \delta p_* = \Gamma_1 p_* \Bigl( -1+(-\alpha/2 +{\cal S}) R_* \Bigr) 
   \delta r/R_*
 \label{eq_slope}
 \end{equation}
With this new boundary condition our eigenvalue problem becomes more
complicated and delicate.  We compute the eigenvalues with a
modified Castor method in which the $\delta p_*$ of
Eq.~\ref{eq_slope} replaces the usual $\delta p_*=0$ boundary
condition.  The presence of strange modes and their frequent
proximity to the regular modes makes the Castor search difficult.
We therefore first compute all the nonadiabatic eigenvalues for the
$\delta p_*=0$ case with a general eigenvalue solver (\eg Glasner \&
Buchler 1993) and then switch on the new boundary condition
gradually.

\centerline{{\vbox{\epsfxsize=10cm\epsfbox{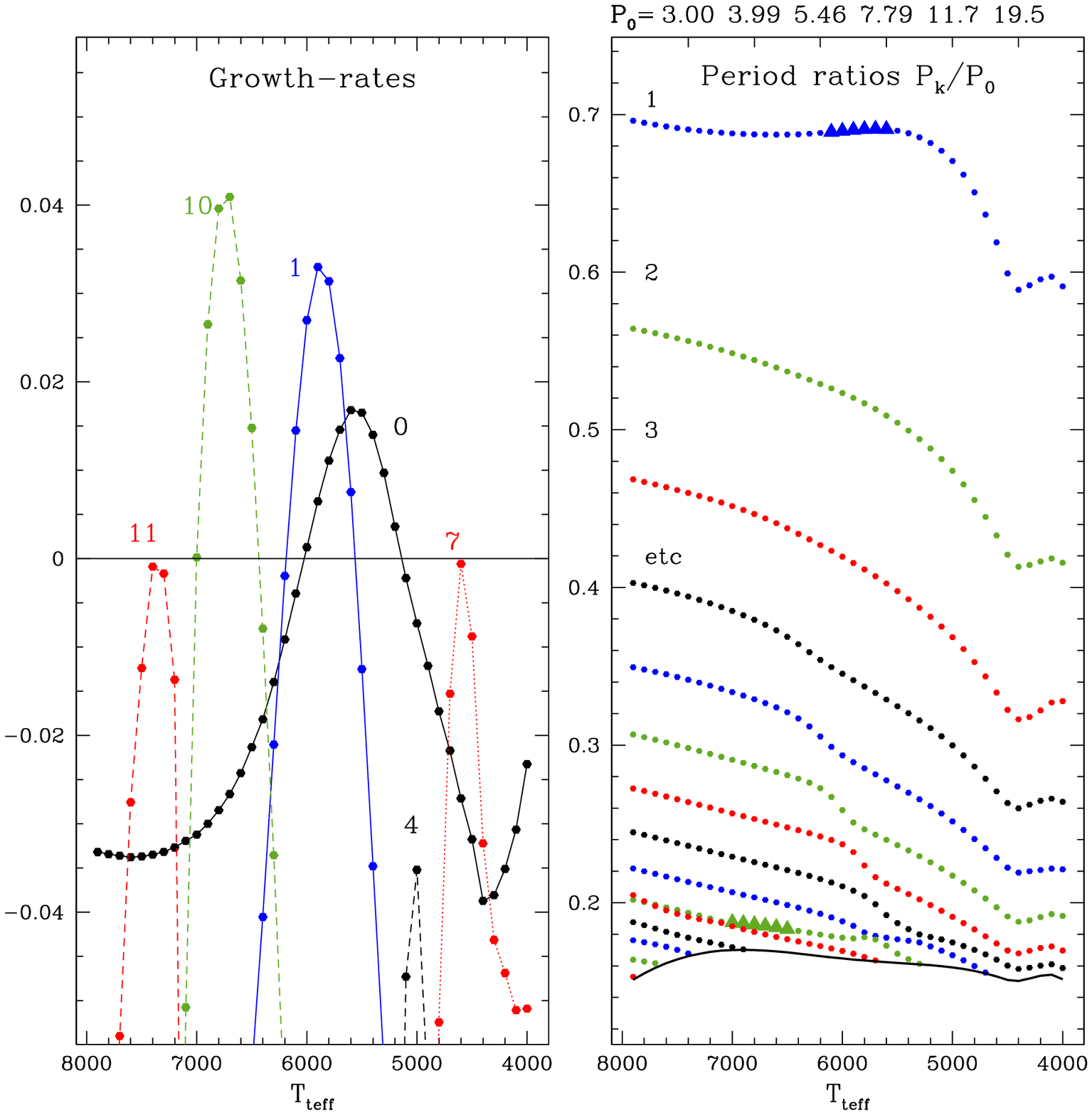}}}}
\noindent{\small Fig.~3. -- Cepheid model, M=6\Mo, L=3000\Lo;
\cf Fig.~2 for legend.}
\vskip 3mm

\section{Turbulent Convective Cepheid Models}

Turbulent convection changes the structure of the outer envelope, as
well as the pulsational driving.  To study its effects on the
strange modes and to ascertain that they survive convection, we have
used the Florida code (Yecko, Koll\'ath \& Buchler 1998) with the
$\alpha$'s defined as in Feuchtinger, Buchler \& Kollath (2000) with
values that are very similar to those used in our study of first
overtone Cepheids, \viz
  $\alpha_d$=  2.177,
  $\alpha_c$= 0.4,
  $\alpha_s$=  0.433,
  $\alpha_\nu$=  0.12,
  $\alpha_t$=  0.001,
  $\alpha_r$=  0.4,
  $\alpha_p$=  0 and
  $\alpha_l$=  1.5.
We use the OPAL opacities of Iglesias \& Rogers (1996) and of
Alexander \& Ferguson (1994).
 
In Figs.~2, 3 and 4 we show the results for 3 sequences of Cepheid
models of fixed mass and luminosity over a broad range of \Teff.
The left subfigures show the growth-rates as a function of the
\Teff\ of the equilibrium models.  The growth-rates for the unstable
modes are positive.  Some modes are very stable and lie below the
graph.

As a function of \Teff\ the modal number (ordering with period) of
the strange mode changes since it switches its properties with the
regular modes through avoided crossings (\cf Paper I).  It is
therefore ambiguous how we should label the modes along a sequence,
but this labelling is not important for us since we are interested
in finding out whether there is an unstable strange mode and what
its period is, both of which are unambiguous.

The instability strip of the 8\Mo\ in Fig.~2 extends from the
fundamental blue edge at \Teff = 5700\K\ to the fundamental red edge
at \Teff = 4600\K.  An unstable strange mode exists from 7700\K\
through 6100\K\, which is the 11th overtone at the lower \Teff, but
which gradually morphs into the 8th overtone by the time it gets to
8000\K.  A correspondingly erratic behavior shows up in the period
ratios.  It is not a numerical artifact, but arises from avoided
level crossings (\cf Paper I) that are also familiar from the
qunatum mechanical behavior of an atom in a B field and from
the nonradial $p + g$ spectrum in unevolved stars.

\centerline{{\vbox{\epsfxsize=10cm\epsfbox{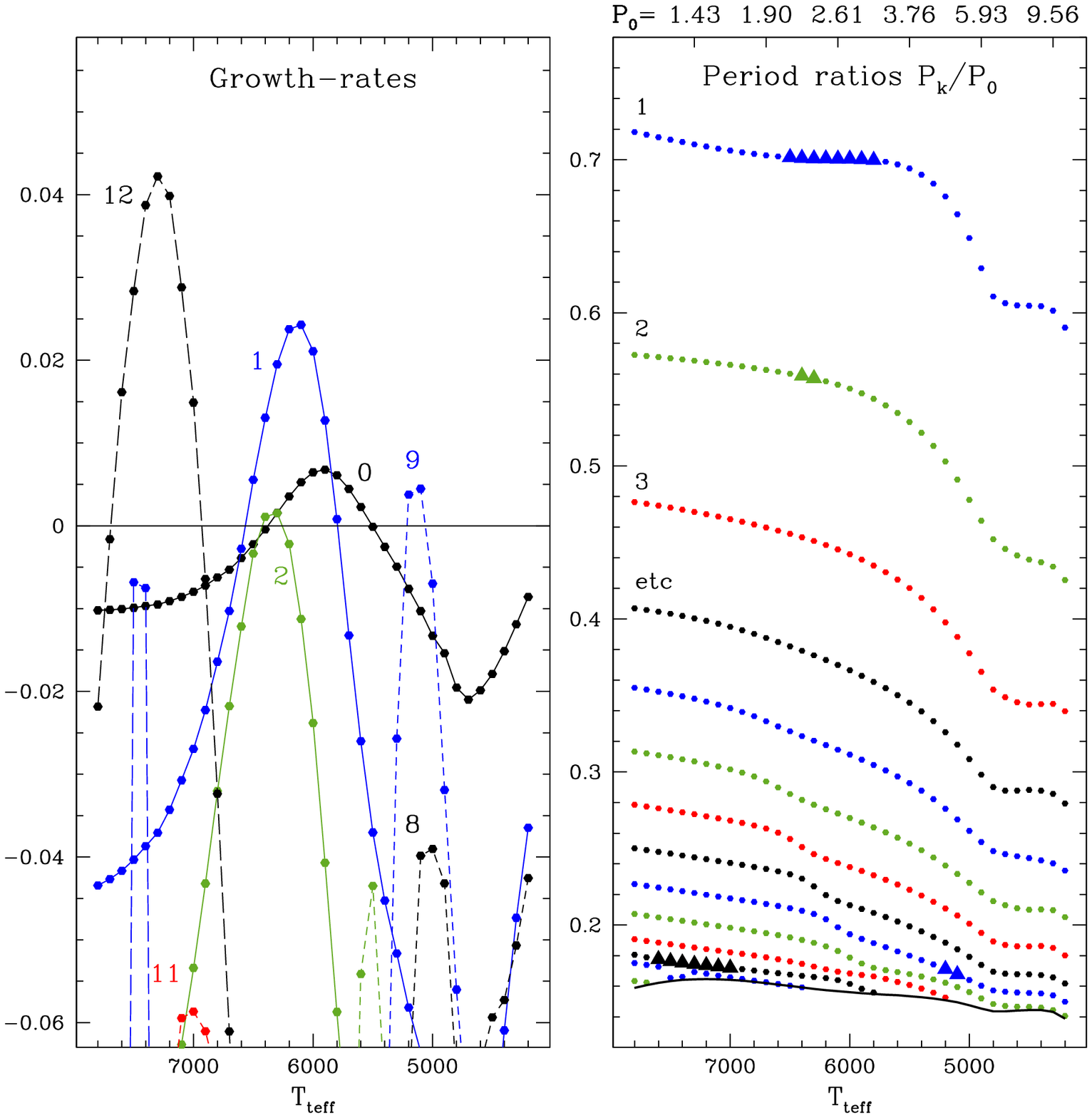}}}}
\noindent{\small  Fig.~4. -- 
Cepheid model, M=4\Mo, L=860\Lo;
\cf Fig.~2 for legend.}
\vskip 4mm

The instability strip of the 6\Mo\ model in Fig.~3 extends from the
first overtone blue edge at \Teff = 6200\K\ to the fundamental red
edge at \Teff = 5150\K.  From 7000\K\ to 6450\K\ the 10th overtone
is unstable.  Around 5000\K\ the 8th overtone just misses being
unstable.

The instability strip of the 4\Mo\ model Fig.~4 extends from the
first overtone blue edge at \Teff = 6550\K\ to the fundamental red
edge at \Teff = 5500\K, and it includes a small region with second
overtone instability.  From 7700\K\ to 6950\K\ the 12th overtone is
unstable, and from 5200\K\ to 5100\K\ it is the 9th overtone.

\section{Turbulent Convective RR Lyrae Models}

We consider an RR Lyrae model with M=0.65\Mo, L=45\Lo\ with X=0.70,
Z=0.02.  Fig.~5 shows that the instability strip of the model
extends from the first overtone blue edge at \Teff = 7250\K\ to the
fundamental red edge at \Teff = 6000\K.  From 8300\K\ to 7200\K\ the
10th overtone is unstable, and from 5900\K\ to 5800\K\ the 8th
overtone is unstable.

Again, these results show that an excited strange mode can exist on
either side of the classical instability strip.

\section{Nonlinear Behavior}

The full amplitude pulsations have been computed with the Florida
hydrodynamics code (Koll\'ath, Beaulieu, Buchler \& Yecko, 1998)
with the same mesh and parameters as used for the linear properties.
The pseudo-viscosity was taken to be zero in our calculations.

\centerline{{\vbox{\epsfxsize=10cm\epsfbox{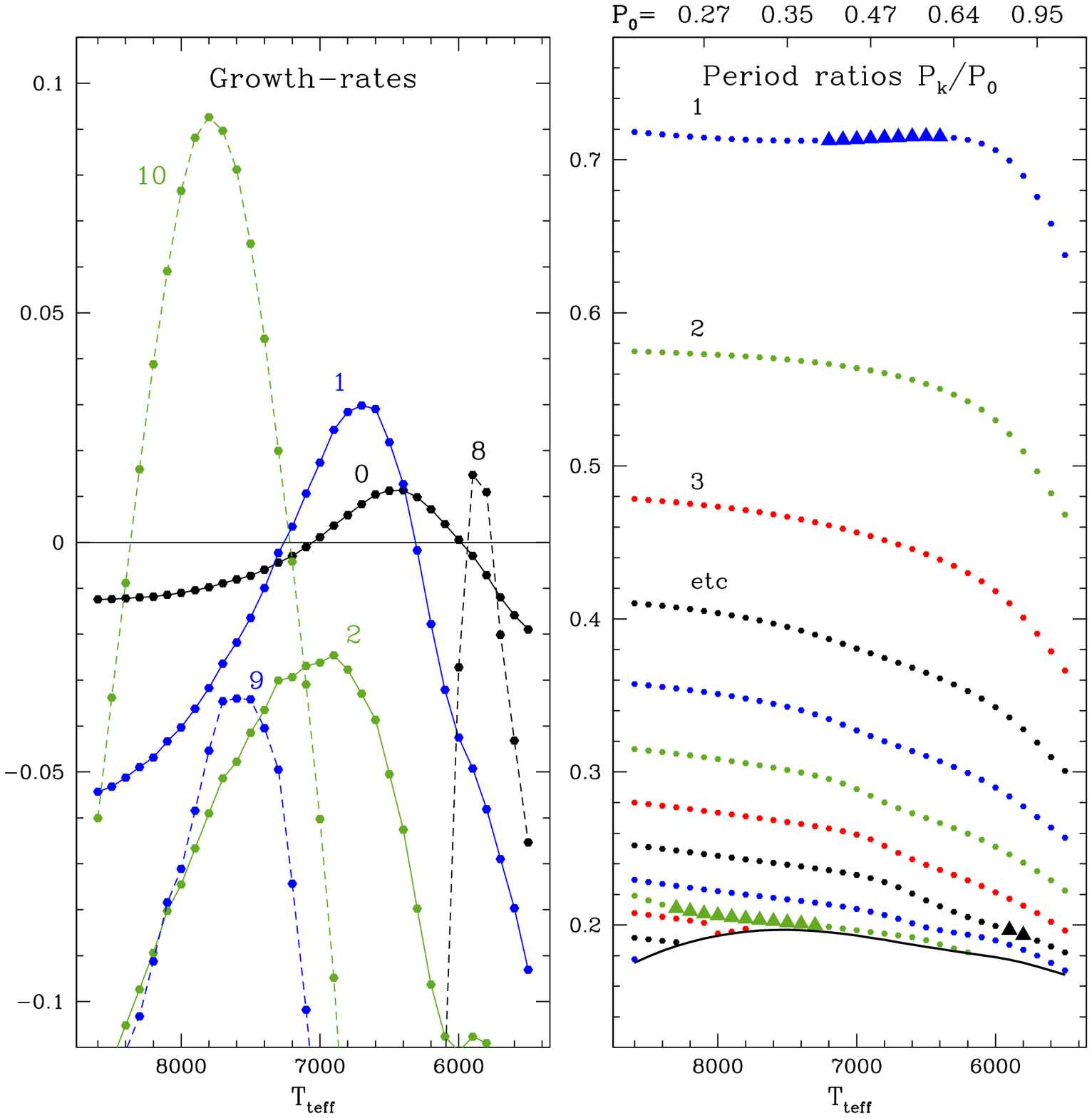}}}}
\noindent{\small Fig.~5. -- 
RR Lyrae model, M=0.65\Mo, L=45\Lo;
\cf Fig.~2 for legend.}
\vskip 3mm

In Fig.~6 we exhibit the light and radial velocity curves for three
selected Cepheid models and on RR Lyrae model. The lightcurves are
measured in millimags, and the radial velocities in units of 100
m/s.  Note that five of the models (1, 3 -- 6) are blue strange
variables, \ie they lie to left of the blue edge, and two (2 and 7)
are red strange variables, \ie they lie to right of the blue edge.
Interestingly they all have about the same pulsation amplitudes.
Because the oscillation amplitude is so small and it is so confined
to the surface region, there is not much matter below the
photosphere that can modulate the heat flow.  The light-curve
amplitudes are therefore also small, in the millimagnitude regime,
but they are observable.

The temporal structure of the photospheric velocities appears
strange at first.  In Fig.~7 we therefore show the behavior of the
velocities throughout most of the model, in fact Lagrangean zones 65
through 145 of a 150 mass-shell model.
The interior zones are essentially quiescent because of the high modal
number of this mode.  On the other hand the amplitude of the very
outer zones increases considerably (but is not relevant because it
contains so little mass and is considerably above the photosphere.  The
photosphere itself, which is shown as a single dashed line, is
relatively quiescent.  For reference we also show the hydrogen
partial ionization region with three adjacent dashed lines. The
helium ionization lies below in zones that are not shown because the
velocities are small.  The photosphere is seen to lie in a 
relatively quiescent region and is subjected to inward and outwards
moving ripples that account for its sometimes strange looking
behavior in Fig.~6.

\centerline{{\vbox{\epsfxsize=11cm\epsfbox{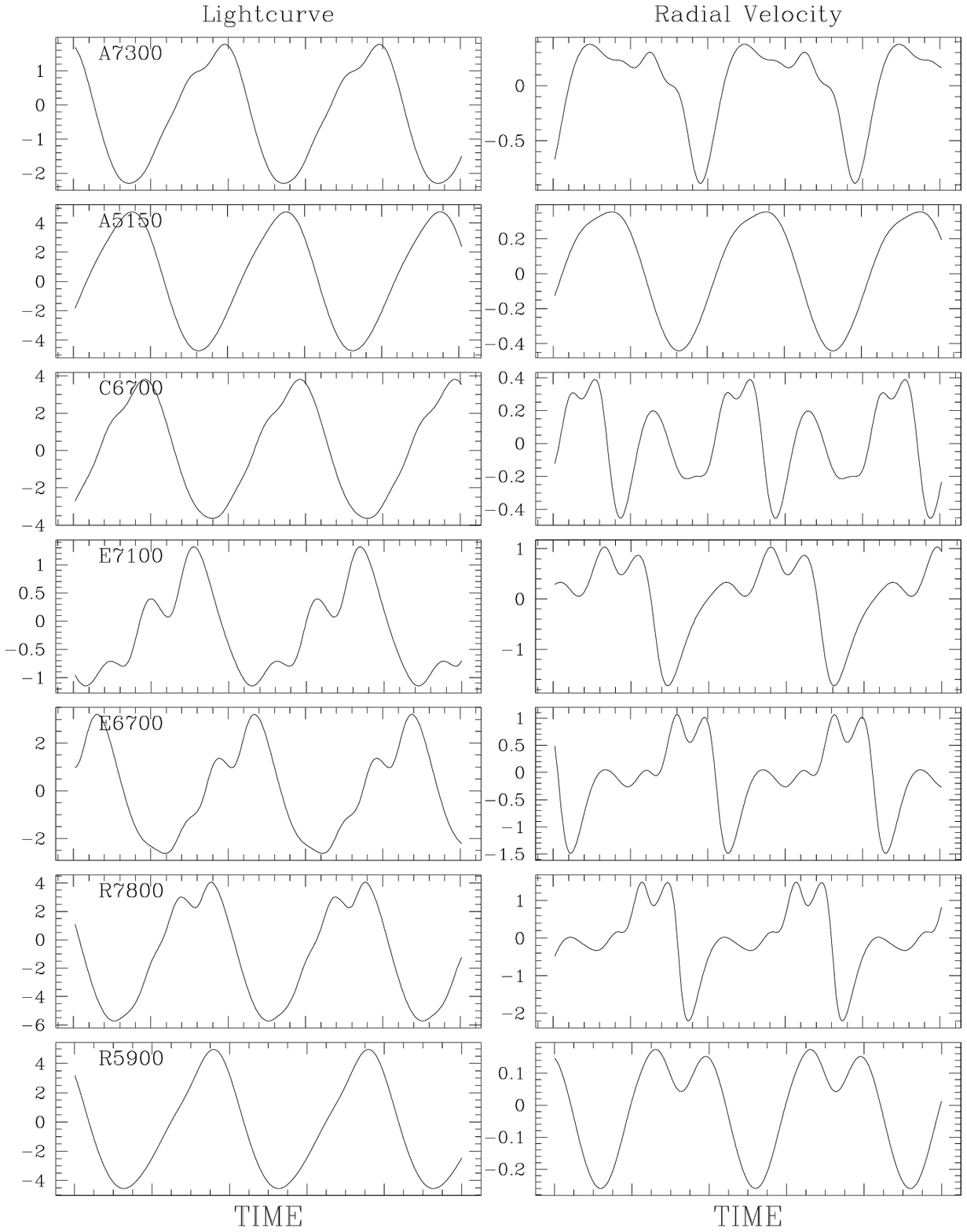}}}}
\noindent{\small Fig.~6. -- 
{\sl Left}: Light curves (in millimag): 
{\sl Right:} radial velocity curves (in 100m/s).
From top to bottom,  
Cepheids: M=4\Mo [A], \Teff =7300 and 5150\K, M=6\Mo [C], \Teff =6700\K, 
M=8 [E], \Teff =7100 and 6700\K, 
RR Lyrae: M=0.65\Mo [R], \Teff=7800 and 5900\K.}
\vskip 3mm

There are two shortcomings to these calculations.  First, the
(linear) special boundary condition at the outer surface
(Eq.~\ref{eq_slope}) has not been incorporated in the nonlinear code,
but we do not expect that to cause a major distortion, for the
pressure is so small at the boundary.  Second, what is perhaps more
severe, there is no provision for acoustic leakage in the nonlinear
hydrodynamics code for waves with frequencies above the isothermal
threshold $\omega_c$.  

The presence of higher frequency reflected waves is perhaps the
reason for the pronounced harmonic wiggles on the light curves and
especially in the radial velocity curves.  Indeed, because spherical
effects are negligible, the higher order strange modes are all
in resonance.  However, if they are excited along with the
lowest strange mode, then the even harmonics should be absent, but
they are not.

\section{Discussion and Conclusion} 

The improved boundary conditions have a small effect on the
behavior of the deep modes, mostly because the pressure variation at the
stellar surface $\delta p_*$ is always small.  However, they do
affect the higher overtones as they approach the cutoff frequency by
shifting their periods, sometimes causing inversions in the modal
ordering and affecting the stability of the modes.

Turbulent convection does not kill the linear instability of the
strange modes, as one might have feared.  On the contrary, they are
as unstable as with the radiative models.  One even finds unstable
strange modes to the right of the fundamental red edge.

The growth-rates of the strange modes on the hot side of the
instability strip are not very sensitive to turbulent convection.
On the other hand convection is very strong in the cooler models,
and the values of the stability coefficients of the strange modes on
the cool side are consequently less certain than they are on the hot
side.

\centerline{{\vbox{\epsfxsize=8cm\epsfbox{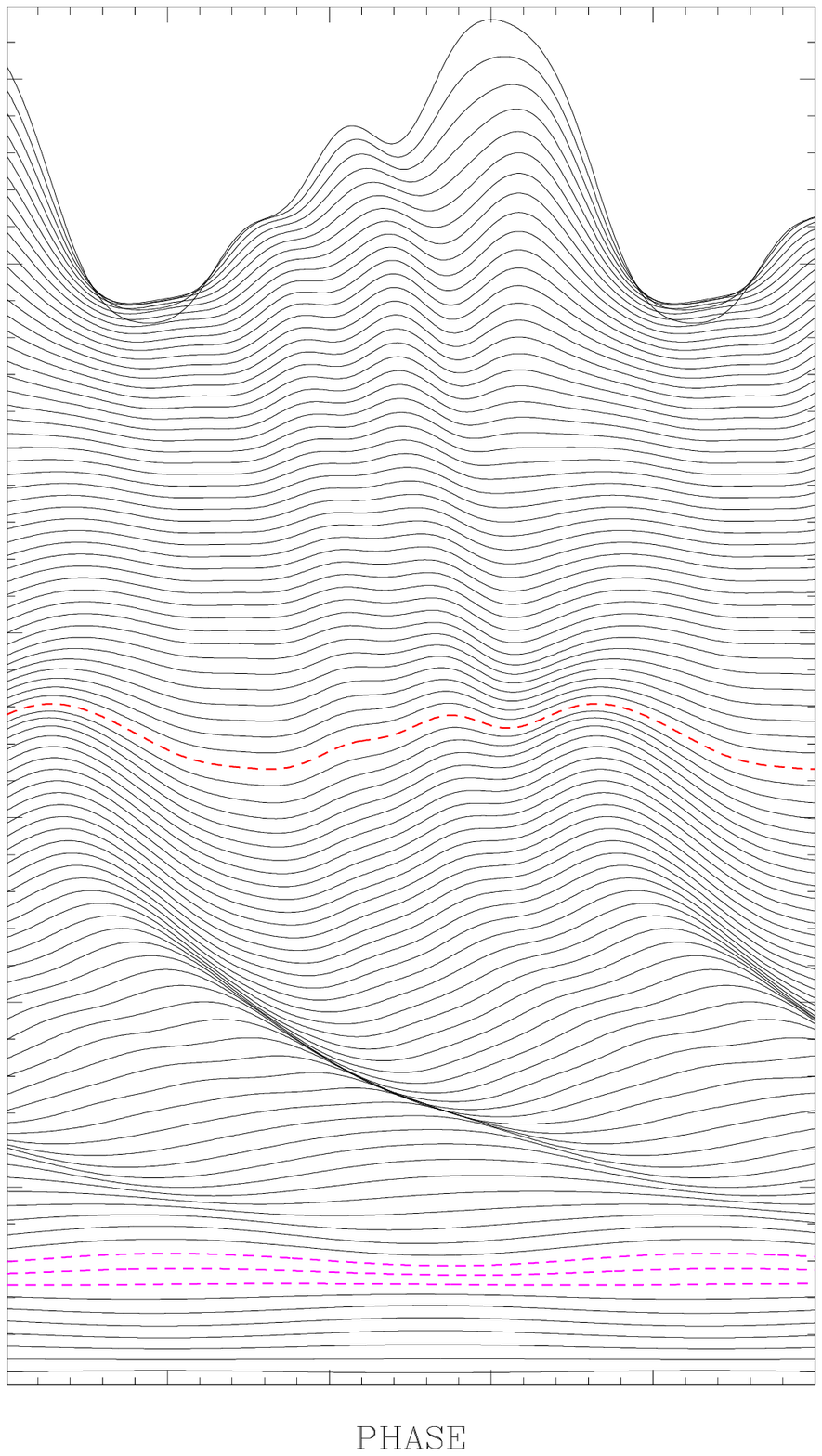}}}}
\noindent{\small Fig.~7. -- 
Radial velocity profile for the Cepheid model  M=6\Mo, \Teff =6700\K;
the individual Lagrangean zones (zones 65 through 145 for a 150 zone
model) are shifted vertically for better visualization;  the dashed
line corresponds to the photospheric velocity. the three adjacent
dashed lines denote the hydrogen partial ionization region}
\vskip 3mm

So, do strange Cepheids really exist in nature?  We find the
occurrence of self-excited strange modes to be quite robust, \ie
largely independent of any modelling parameters.  For an individual
model the change in modelling parameters may have a large effect,
but when one considers a whole sequence of models on finds that
modelling parameters (\eg the fineness of the grid) merely cause a
small shift (\eg in \Teff).  Especially in the high luminosity
Cepheid models, the strange mode has a period well above the
isothermal cutoff period as the right hand side of Fig.~2 shows.
For the low mass Cepheids and the RR Lyrae some of the unstable
strange modes occur very close to the isothermal cutoff frequency
$\omega_c$.  This cutoff frequency however is somewhat uncertain
because it depends on the structure of the dilute atmosphere which
itself is sensitive to the treatment of radiative transport (treated
here in the customary diffusion approximation).

We find furthermore that the full amplitude (nonlinear) pulsations
are quite robust in all cases.  However, the values of the pulsation
amplitudes are somewhat uncertain because they depend on modelling
parameters as they did for the classical Cepheid pulsations.  In
particular the turbulent eddy viscosity has a large effect.  (We
recall that our convective parameters $\alpha$'s were calibrated to
give good agreement with the observations of classical Cepheids,
Feuchtinger \etal 2000).

The detection of this type of pulsation would be very important
because it would add very strong additional constraints on the
Cepheid and RR Lyrae modelling.  The search for millimagnitude
variability on either side of the classical instability strip with
periods approximately 1/4 to 1/5th the fundamental period (\cf
Figs.~1 --3, 5) would be required.  Hopefully this will become
possible with COROT.

 \section{Acknowledgements}
 It is a pleasure to thank Phil Yecko and Bernard Whiting 
for fruitful discussions.
This work has been supported by NSF (AST 98-19608).

\end{document}